\documentclass[iop]{emulateapj}

\usepackage{amsmath}
\usepackage{natbib}

\slugcomment{}

\newcommand{\be}{\begin{equation}}
\newcommand{\ee}{\end{equation}}

\shorttitle{Non-circular orbits}
\shortauthors{L\'opez-Corredoira \& Gonz\'alez-Ferna\'andez}

\begin{document}

\title{Radial motions in disk stars: ellipticity or secular flows?}
\author{M.  L\'opez-Corredoira\altaffilmark{1,2} and C. Gonz\'alez-Fern\'andez\altaffilmark{3}}
\altaffiltext{1}{Instituto de Astrofisica de Canarias, E-38205 La Laguna, Tenerife, Spain; martinlc@iac.es}
\altaffiltext{2}{Departamento de Astrofisica, Universidad de La Laguna, E-38206 La Laguna, Tenerife, Spain}
\altaffiltext{3}{Institute of Astronomy, University of Cambridge, Madingley Road, Cambridge, CB3 0HA, UK}

\begin{abstract}
Average stellar orbits of the Galactic disk may have
some small intrinsic ellipticity which breaks the exact axisymmetry and
there may also be some migration of stars inwards or outwards. Both phenomena can be detected
through kinematic analyses.
We use the red clump stars selected spectroscopically from APOGEE (APO Galactic Evolution Experiment), with known distances and radial velocities, to measure the radial component of the Galactocentric velocities within 5 kpc$<R<$16 kpc, $|b|<5^\circ$ and within 20 degrees from
the Sun-Galactic center line. The average Galactocentric radial velocity is
$V_R=(1.48\pm 0.35)[R({\rm kpc})-(8.8\pm 2.7)]$ km/s outwards in the explored range,
with a higher contribution from stars below the Galactic plane.
Two possible explanations can be given for this result: i) the mean orbit of the disk stars is intrinsically elliptical with a Galactocentric radial gradient of eccentricity around 0.01 kpc$^{-1}$; or ii) there is a net secular expansion of the disk, in which stars within $R\approx 9-11$ kpc are migrating to the region $R\gtrsim 11$ kpc at the rate of $\sim 2$ M$_\odot $/yr, and stars with $R\lesssim 9$ kpc are falling toward the center of the Galaxy. This migration ratio would be unattainable for a long time and it should decelerate, otherwise the Galaxy would fade away in around 1 Gyr. At present, both hypotheses are speculative and one would need data on the Galactocentric radial velocities for other azimuths different to the center or anticenter in order to confirm one of the scenarios.
\end{abstract}


\section{Introduction}

The disk of spiral galaxies like our Milky Way is represented approximately by
a stationary axisymmetric component. However, the exact representation of
disks may depart from this simple first-order picture. Average stellar orbits might have
some small intrinsic ellipticity which breaks the exact axisymmetry and the orbits
may be stationary, like for a planet of the solar system 
which always returns to the same point with respect to the Sun, or
there may be some long-term evolution such as for instance a possible migration of stars inwards or outwards. 

Most disks exhibit a wealth of non-axisymmetric structures
(Rix \& Zaritsky 1995); about one third of them are substantially lopsided at a 2.5 disk exponential scale length, although the spiral pattern couples significantly to the estimate of the intrinsic ellipticity and their measurement may represent an upper limit on the true potential triaxiality. Lopsidedness is quite typical in disk galaxies and this may be interpreted as a pattern of elliptical orbits (Baldwin et al. 1980; Song et al. 1983).
Non-circular streaming motions were also observed in the gas motions
of other galaxies (Sellwood \& Zanmar S\'anchez 2010); however, the stellar kinematics is
usually more regular and symmetric than the gas kinematics (Pizzella et al. 2008). Therefore,
further research is needed for the stellar population and, in particular, this can be more accurately analyzed in the Milky Way.

In our Galaxy, Siebert et al. (2011) and Williams et al. (2013) were able to obtain, from RAVE (Radial Velocity Experiment) spectroscopic data, a significant measurement of a Galactocentric radial velocity gradient outwards. This gradient was measured within -2 kpc$<(R-R_\odot )<+1$ kpc and mostly measures local streaming motions. A large-scale feature of the Galactic disc should be observed over a wider range of Galactocentric distances and this is what we will carry out here, thanks to the higher depth of the APOGEE (APO Galactic Evolution Experiment) survey (see \S \ref{.redclump}). The analysis for -3 kpc$<(R-R_\odot )<+8$ kpc along the Sun-Galactic center line is carried out in \S \ref{.proper} and \S \ref{.results}.

The interpretation of this gradient of radial velocity can be given in terms of
elliptical orbits. In \S \ref{.ellipt}, we statistically constrain their ellipticities
only with the radial velocities along the Sun-Galactic center line.

Another tentative explanation is the existence of a net migration of stars outwards.
Stellar radial migrations have indeed been a useful hypothesis to explain breaks in surface brightness 
(S\'anchez-Bl\'azquez et al. 2009), 
the formation of the thick disk (Sales et al. 2009), metallicity distributions (Grand et al.
2015), etc., and it is indeed theoretically expected as a consequence of
the resonances of the bar and transient spiral arms in the disk (Halle et al. 2014:
Roskar \& Debattista 2015). 
Here we will derive the ratio for the necessary average migration to produce our
observed Galactocentric radial velocity gradient (\S \ref{.migration}).

\section{Data from APOGEE: selection of red clump giants}
\label{.redclump}

APOGEE is an H-band high-resolution spectroscopic survey (Eisenstein et al. 2011) of the third stage of the Sloan Digital Sky Survey project (SDSS-III; Gunn et al. 2006). A detailed description of the target selection and data reduction pipeline is presented in Zasowski et al. (2013). For the purposes of this paper, we use the spectroscopically selected sample of red clump giants (RCGs) presented by Bovy et al (2014), updated for the Data Release 12 (DR12) of SDSS-III.

The narrow luminosity function distribution of RCGs (Castellani et al. 1992) makes them very appropriate standard candles that trace the old stellar population of the Galaxy.
The catalog derived from APOGEE-DR12 contains a total of 19\,937 RCGs (with the constraint of the parameter ADDL\_LOGG\_CUT=1, which includes only the sources with the extra constraint in $\log g$ of Eq. (9) in Bovy et al., which makes the selection more accurate) for which the distance is determined using $M_K=-1.61$ and accurate measurements of their radial motions are provided. The narrowness of the RCG locus in color--metallicity--luminosity space allows distances to the stars to be assigned with an accuracy of 5\%-10\%. The purity is estimated to be about 95\% ; we neglect the
possible systematic errors due to the possible errors in the distances of these stars.
 As explained in the next section, we only take the in-plane regions near the anticentre, we constrain our sample within $|b|<5^\circ$ and $|\ell-180^\circ |<20^\circ $ or 
(or $|\ell |<20^\circ $ and $R>5$ kpc), so we work with a sample of 3\,160 RCGs. 

\section{Deriving Galactocentric radial velocities from heliocentric radial velocities}
\label{.proper}

\begin{figure}
\centering
\includegraphics[width=7.cm]{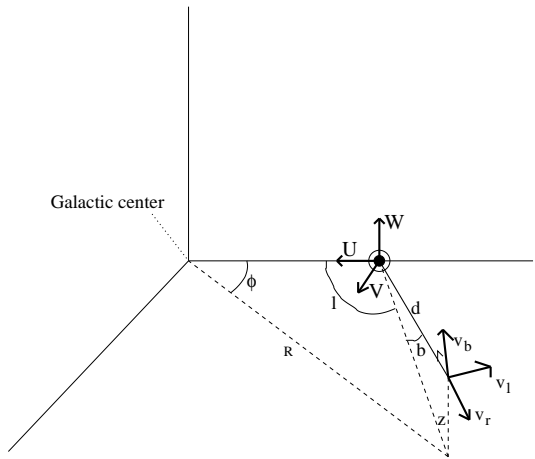}
\caption{Sketch of the kinematics of a star with respect to the Sun.}
\vspace{.2cm}
\label{Fig:convvel}
\end{figure}

The 3D velocity of the combination of radial velocity ($v_r$) and 
tangential velocities ($v_\ell $, $v_b$) is related to the velocity in the reference system $U$, $V$, $W$ as plotted in Fig. \ref{Fig:convvel} by
\begin{equation}
\label{velhel}
v_r=U_*\cos \ell \cos b+V_*\sin \ell \cos b+W_*\sin b
,\end{equation}\[
v_\ell =-U_*\sin \ell+V_*\cos \ell
\]\[
v_b=-U_*\cos \ell \sin b-V_*\sin \ell \sin b+W_*\cos b
,\]
where $(U_*, V_*, W_*)$ is the velocity of a star relative to the Sun in the system $(U,V,W)$.

A star with Galactocentric distance $R$, azimuth $\phi $ and vertical distance from the plane $z$ will have a Galactocentric velocity with a radial component ($V_R$; we define it as positive outwards), an azimuthal component (rotation speed $V _\phi$) and a vertical motion ($W$) which are related to the heliocentric velocities by 
\begin{equation}
\label{velgal}
U_*=-U_\odot +V _\phi \sin \phi -V_R\cos \phi 
,\end{equation}\[
V_*=-V_{g,\odot }+V _\phi \cos \phi +V_R\sin \phi 
\]\[
W_*=-W_\odot +W
,\]
where $V_{g,\odot }=V _\phi (R_\odot ,z=0)+V_\odot $ and 
$(U_\odot , V_\odot , W_\odot )$ is the velocity of the Sun with respect to the local 
standard of rest (LSR).
Here, we adopt the values $R_\odot =8$ kpc; $V_{g,\odot }=244\pm 10$ km/s, $U_\odot =10\pm 1$ km/s, 
$V_\odot =26\pm 3$ km/s (Bovy et al. 2012), and $W_\odot =7.2\pm 0.4$ km/s (Sch\"onrich et al. 2010).

See L\'opez-Corredoira (2014) and 
L\'opez-Corredoira et al. (2014) for the derivation of $V_\phi $ and $W$ respectively from the tangential velocities. To obtain $V_R$ without any assumption on $V_\phi $ or $W$, 
in principle, by joining Eqs. (\ref{velhel}) and (\ref{velgal})
we could use both tangential velocities and radial velocities of the stars through
\begin{equation}
V_R=v_\ell \sin(\phi +\ell )-v_r\cos(\phi +\ell)\sec b-U_\odot \cos \phi
\end{equation}\[
\ \ \ \ \ \ +V_{g,\odot }\sin \phi+(W_\odot -W)\cos (\phi +\ell)\tan b
,\]
but we have checked that the tangential velocities in the APOGEE stars given by Bovy et al. (2014) have got very large error bars and introduce much noise in our calculations. Therefore, we use only the information of the radial heliocentric velocities through the expression also derived
from Eqs. (\ref{velhel}) and (\ref{velgal}):
\begin{equation}
\label{VR}
V_R=-\frac{v_r}{\cos (\phi +\ell )\cos b}-\frac{\cos \ell}{\cos (\phi +\ell)}U_\odot -
\frac{\sin \ell}{\cos (\phi+\ell )}V_{g,\odot }
\end{equation}\[
\ \ \ \ \ \ +\frac{\tan b}{\cos (\phi +\ell)}
(W-W_\odot )+\tan (\phi +\ell )V_\phi
.\]
This is the relationship that we will use throughout this paper. 
The disadvantage of Eq. (\ref{VR}) is that
we are model dependent, since we need to know the values of $V_\phi$ and $W$, but we can make 
an appropriate selection of regions in which this dependence is very small. In particular,
we choose $|b|<5^\circ$, which makes negligible the contribution of $W$ or $W_\odot $ and 
$|\ell-180^\circ |<20^\circ $ (or $|\ell |<20^\circ $ and $R>5$ kpc), in order to avoid
low values of $\cos (\phi+\ell )$ and to reduce the impact of the error in $V_\phi$.
We set $W=0$, neglecting the vertical motions (which might, however, be substantial (L\'opez-Corredoira et al. 2014), but in any case without a significant contribution here given the
low values of $b$). The rotation speed is taken from Bovy et al. (2012), who also
use APOGEE data to derive it. We have performed tests with other rotation curves and 
we do not obtain different results within the error bars. There may be some gradient of
the rotation speed with $z$ and this is apparently quite conspicuous at $R\gtrsim 14$ kpc 
(L\'opez-Corredoira 2014), but since we are using low $b$ data, this can be neglected.

The velocity $V_R$ might be related to the Oort constants (Siebert et al. 2011, Famaey et al. 2012) but this is only valid within the approximation of a Taylor expansion within the solar
neighborhood, for heliocentric distances $\lesssim 1$ kpc. Since we are going much
farther away, we will not derive these constants.

\section{Results}
\label{.results}

\begin{figure}
\vspace{1cm}
\centering
\includegraphics[width=8.cm]{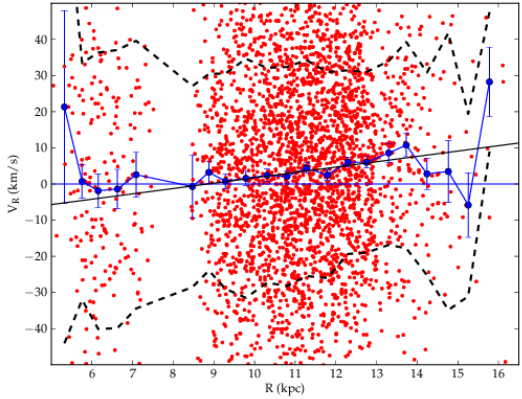}
\caption{Radial galactocentric velocity derived from Eq. (\protect{\ref{VR}}) with radial
heliocentric velocities from APOGEE for RCG sources 
within a region close to the Galactic center-Sun line. The blue line and its error bars
represent the average within bins of $\Delta R=0.5$ kpc.
The region between both dashed lines is the zone within one rms of dispersion of the points.}
\vspace{.2cm}
\label{Fig:vrvsr}
\end{figure}

Fig. \ref{Fig:vrvsr} shows the measured values of $V_R$ as a function of $R$, in which we have grouped data in bins of $\Delta R=0.5$ kpc and we have carried out a linear weighted fit that gives the result:
\begin{equation}
\label{VRfit}
V_R({\rm km/s})=(1.48\pm 0.27)\times [R({\rm kpc})-(8.84\pm 0.45)]
.\end{equation}
The slope is $>5\sigma$ away from zero, which means that we significantly detect a variation of $V_R$ with $R$. Given our constraints, this applies to the sources which are close to the Galactic center-Sun line. 
The error bar of mean radial galactocentric velocity is inversely proportional to the root square of the number of points in each bin and proportional to the rms of the velocities. Certainly, for a much smaller number of stars the error bar would be larger and it would reduce the significance of our detection, but we have found a detection of a positive slope with over 5 sigma of significance, so clearly we are using a sufficient number of stars for our purposes.
The quantity 0.27 corresponds only to the
error of the fit, and we should also take into account the error in the used parameters, derived from an error expansion of Eq. (\ref{VR}), neglecting the possible covariance terms if there were some correlations among the parameters:
\begin{equation}
\left({\rm Error}\left[\frac{dV_R}{dR}\right]\right)^2=0.27^2
\end{equation}\[
+\left(\left(\frac{d\left\langle\frac{\cos \ell}
{\cos (\phi +\ell)}\right\rangle}{dR}\right){\rm Error}[U_\odot]\right)^2
\]\[
+\left(\left(\frac{d\left\langle\frac{\sin \ell}
{\cos (\phi +\ell)}\right\rangle}{dR}\right){\rm Error}[V_{g,\odot}]\right)^2
\]\[
+\left(\left(\frac{d\left\langle\frac{\tan b}
{\cos (\phi +\ell)}\right\rangle}{dR}\right){\rm Error}[W-W_\odot]\right)^2
\]\[
+\left(\left(\frac{d\left\langle\tan (\phi+\ell)
\right\rangle}{dR}\right){\rm Error}[V_\phi]\right)^2
\]\[
+\left(\left\langle\tan (\phi+\ell)
\right\rangle \left(\frac{d{\rm Error}[V_\phi]}{dR}\right)\right)^2
.\]
Because of the constraints in the selected region, the derivatives of the
brackets are small:
$\frac{d\left\langle\frac{\cos \ell}
{\cos (\phi +\ell)}\right\rangle}{dR}$=-0.0023 kpc$^{-1}$,
$\frac{d\left\langle\frac{\sin \ell}
{\cos (\phi +\ell)}\right\rangle}{dR}$=-0.022 kpc$^{-1}$,
$\frac{d\left\langle\frac{\tan b}
{\cos (\phi +\ell)}\right\rangle}{dR}$=0.00066 kpc$^{-1}$, and
$\frac{d\left\langle \tan (\phi+\ell) \right\rangle}{dR}$=-0.017 kpc$^{-1}$.
With these numbers and the errors of the parameter given above, we get that 
${\rm Error}\left[\frac{dV_R}{dR}\right]=0.35$ km/s/kpc, somewhat larger than the 
previous 0.27 with just the fitting error. In any case, we can claim that
the slope is $\gtrsim 4\sigma$ away from zero.

However, the value of $R$ which gives exactly $V_R=0$, as expected in circular orbits,
is not so well determined: the fit gives $R_0=8.84\pm 0.45$ kpc as the zero point,
but we must add to this error that of $U_\odot $, which dominates the uncertainty in this measurement. Bovy et al. (2012) is quite optimistic to give an error of only 1 km/s
to this quantity, whereas some other authors like Sch\"onrich (2012) derive a value +4 km/s higher than the one by Bovy et al. (2012). If we take the ${\rm Error}[U_\odot]=4$ km/s, given that $\left\langle\frac{\cos \ell}{\cos (\phi +\ell)}\right\rangle \approx 1.0$ in our sample, we obtain ${\rm Error}[R_0]=2.74$ kpc instead of only 0.45 kpc. That is, $R_0=8.84\pm 2.74$ kpc. 
Of course, the value of $R_0$ will also change if we change $R_\odot $; for instance, increasing $R_\odot $ to 8.5 kpc instead of 8.0 will increase the value of $R_0$ in $\approx 0.5$ kpc.

The trend in $\frac{dV_R}{dR}$ was also observed by Siebert et al. (2011): $\left \langle \frac{dV_R}{dR}\right \rangle=+4$ km/s/kpc over -2 kpc$<(R-R_\odot )<+1$ kpc. Furthermore,
Williams et al. (2013) found that the gradient in that range is marked below the plane 8 km/s/kpc for negative latitudes and vanishing to zero above the plane, with a $z$-gradient thus also present. If we perform the same analysis as Williams et al., but for -3 kpc$<(R-R_\odot )<+8$ kpc, 
we also obtain a higher gradient for $b<0$ stars than $b\ge 0$ stars: $2.12\pm 0.77$ km/s/kpc and $0.36\pm 0.68$ km/s/kpc, respectively. Fig. 3 of Siebert et al. (2011) or Fig. 16 of Williams et al. (2013) are comparable to our Fig. \ref{Fig:vrvsr}.
Bovy et al. (2015) also found evidence for non-circular motions from the power spectrum of the velocity fluctuations after subtracting an axisymmetric model.

Dynamical reasons for this gradient will not be explored in this paper, but we will
just interpret the kinematics without relating it to any theoretical model.
There are in the literature some attempts to explain these kinds of observations:
no gradient in the radial velocity could be produced by a bar (Monari et al. 2014), but it can be produced by spiral arms (Faure et al. 2014), for instance.
For our present data, the effect of a net increasing Galactocentric radial velocity over a range of $\sim 10$ kpc cannot be explained with a bias in which we see more or better the stars on the closer side of the spiral arm, which pull them toward the arm, and we do not see as well the stars behind the spiral arms that should provide the expected symmetric view. In such a case we would see fluctuations around zero in the positions where we cross a spiral arm rather than a continuous increase of the velocity as shown in Fig. \ref{Fig:vrvsr}.
Obtaining a higher gradient from $b<0$ than from $b>0$ seems to be in favor of rather a local process, maybe a dominant local stream driving this impression of an outward radial displacement, but again this effect cannot be
local since it is observed along a wide range of $\sim 10$ kpc. Whatever it
is the cause of the present effect, it is not something restricted to a
local event in some place of the Galactic disk, but a large-scale effect.

\section{Interpretation}

The detection of $V_R\ne 0$ means that the mean orbit of the stars is not perfectly circular, either because it is Keplerian elliptical or because there is a component of secular expansion of the disk associated to migration. In this section we explore further the consequences of both hypotheses, bearing in mind that with the data at hand we cannot favor one or the other.

\subsection{Elliptical orbits}
\label{.ellipt}

We want to derive the properties of the mean orbit. Note that the ellipticity of the 
mean orbit is not the same thing that the mean ellipticity of the orbits for individual stars.

For an elliptical orbit, the radial Galactocentric velocity is related to the eccentricity, $e$, through
\begin{equation}
V_R^2=\frac{Ke^2}{R}\frac{\sin^2(\phi-\phi _0)}{1+e\cos (\phi -\phi _0)}
,\end{equation}
where $K=R\,V_\phi ^2$ in circular orbits. For very low eccentricities ($e\ll 1$)
\begin{equation}
\label{VReapp}
V_R\approx V_\phi\,e\sin(\phi-\phi _0)
,\end{equation}
\begin{equation}
\label{e}
e\approx \sqrt{\frac{1}{V_\phi ^2}\left[V_R^2+
\left(\frac{\partial V_R}{\partial \phi}\right)^2\right]}\ge \frac{V_R}{V_\phi}
.\end{equation}

We have no information on the dependence of $V_R$ on $\phi $, so we cannot derive
$\frac{\partial V_R}{\partial \phi}$ and we cannot derive the exact value of $e$
from Eq. (\ref{e}). However, we can evaluate the most likely value: in almost circular orbits, the probability to obtain a value of $\phi $ between $\phi _1$ and $\phi _1+\Delta \phi $ is proportional to the time in this range of azimuths, which is proportional to $\Delta \phi \approx
\Delta \left[\sin ^{-1}\left(\frac{V_R}{V_\phi \,e}\right)\right]$ [from Eq. (\ref{VReapp})]. 
Therefore, the normalized probability to have an eccentricity between $e$ and $e+de$ 
for $\frac{V_R}{V_\phi}\le e\le 1$ is
\begin{equation}
P(e)de\approx \frac{1}{\frac{\pi }{2}-\sin ^{-1}\left(\frac{V_R}{V_\phi }\right)}
\frac{|V_R|}{V_\phi e^2}\frac{de}{\sqrt{1-\frac{V_R^2}{V_\phi ^2e^2}}}
.\end{equation}
Note that this approximation is only correct for low values of $e$ and not for high values close to unity, but since the probability $P(e)$ goes down very fast for high values of $e$ and low
values of $V_R$, it does not matter for our calculations: the important thing is the
localization of the peak of the probability and the errors in the
tail of the distribution for high $e$ do not have an effect. 
Neglecting the term $\sin ^{-1}\left(\frac{V_R}{V_\phi }\right)$ with respect to
$\pi /2$ for $V_R\ll V_\phi $ and convolving
with the Gaussian distribution of values for a given average value of velocity $V_R$ and
corresponding rms $\sigma $, we obtain
\begin{equation}
\label{pe}
P(e)de\approx \frac{2^{1/2}de}{\pi^{3/2}\sigma \,V_\phi \,e^2}
\end{equation}\[\times
\int _{-V_\phi \,e}^{V_\phi \,e}
dx\frac{|x|}{\sqrt{1-\frac{x^2}{V_\phi ^2e^2}}}exp\left[-\frac{(x-V_R)^2}{2\sigma ^2}\right]
.\]

\begin{figure}
\vspace{1cm}
\centering
\includegraphics[width=7.cm]{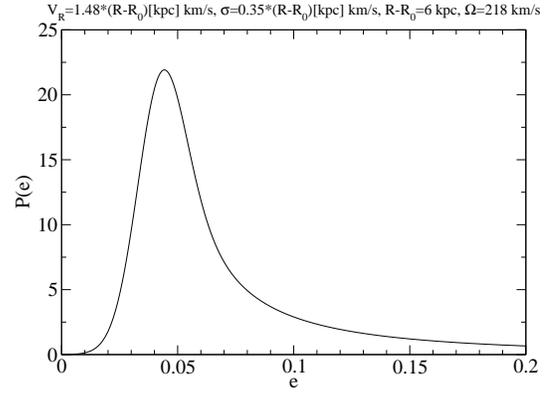}
\caption{Probability distribution of the eccentricity 
given by Eq. (\ref{pe}) for the specified set of parameters.}
\vspace{.2cm}
\label{Fig:prob_e}
\end{figure}

We apply Eq. (\ref{pe}) for the values derived from Sect. \ref{.results}:
$V_R=(1.48\ {\rm km/s/kpc})(R-R_0)$, $\sigma=(0.35\ {\rm km/s/kpc})(R-R_0)$,
and for a constant rotation speed of $V_\phi =218$ km/s. In Fig. \ref{Fig:prob_e}
we show the probability distribution for the case $R-R_0$=6 kpc. The linear fit of
$V_R$ of Sect. \ref{.results} can be translated into
\begin{equation}
e\approx \left[0.009^{+0.014}_{-0.003}(68\% \ {\rm C.L.})\ 
^{+0.077}_{-0.005}(95\% \ {\rm C.L.}) \ {\rm kpc}^{-1}\right]
\end{equation}\[
\times |R-R_0|
.\]
The error distribution is very different from a Gaussian one (see Fig. \ref{Fig:prob_e}).
With the available information we cannot ascertain the position of the major
axis of these ellipses; we would need to explore all values of $\phi $ but we are constrained
at $\phi \sim 0$ for the given reasons. 

\subsection{Secular expansion}
\label{.migration}

Let us assume that the mean orbit is a spiral [$V_R\ne V_R(\phi)$] and Eq. (\ref{VRfit}) represents the average velocity of the whole collection of stars, that $V_R$ in the anticenter direction represents the mean $V_R$ at any azimuth. We admit that this a first--order approach, useful to discuss the order of magnitude of the event, and a more accurate result should take into account the possible dependence with azimuth. 
Then, we have an average outwards motion of 
$V_{\rm rm}(R)=(1.51\pm 0.36)\times [R({\rm kpc})-(8.84\pm 2.74)]$ kpc Gyr$^{-1}$, including the different sources of errors discussed in Sect. \ref{.results}.

The relative variation of stellar mass of a ring with radii between $R$ and $R+dR$ is \begin{equation}
\frac{\dot{M}}{M}(R)=\frac{-1}{R\sigma (R)}\frac{d[V_{\rm rm}(R)R\sigma (R)]}{dR}
,\end{equation} 
where $\sigma (R)$ is the stellar surface density, 
assuming a constant average mass/luminosity ratio throughout the disk. 
With an exponential disk $\sigma (R)\propto e^{-R/h_R}$ and the above 
value of $V_{\rm rm}(R)$, we obtain that the relative gain of stellar mass in a 
ring of radius $R$ is 
\begin{equation}
\label{mdot}
\frac{\dot{M}}{M}(R)=V_{\rm rm}(R)\left(\frac{1}{h_r}-\frac{1}{R}\right)-\frac{dV_{\rm rm}(R)}{dR}
\end{equation}\[
=(1.51\pm 0.36)\times \frac{R^2-R(R_0+2h_R)+R_0h_R}{Rh_R}\ \ {\rm Gyr}^{-1}
.\]
This means that the rings with $R>R_1=\frac{1}{2}R_0+h_R+\sqrt{R_0^2/4+h_R^2}$ kpc are gaining mass whereas the rings with $\frac{1}{2}R_0+h_R-\sqrt{R_0^2/4+h_R^2}<R<\frac{1}{2}R_0+h_R+\sqrt{R_0^2/4+h_R^2}$ are loosing mass, and the rings with $R<\frac{1}{2}R_0+h_R-\sqrt{R_0^2/4+h_R^2}$ are gaining mass again.

Integrating Eq. (\ref{mdot}) for the whole
disk at the region which is gaining mass ($R>R_1$), 
and assuming $h_R=2.0\pm 0.4$ kpc for a thin disk (L\'opez-Corredoira \& Molg\'o 2014) and a local stellar surface density of 
$\sigma _\odot = (3.8\pm 0.4)\times 10^7\,M_\odot/$kpc$^2$ (Bovy \& Rix 2013), we obtain
$\dot{M}(R>R_1)=1.9^{+3.9}_{-1.5}$ M$_\odot $/yr. The linear fit of $V_R$ in Eq. 
(\ref{VRfit}) might not be extrapolated to an infinite Galactocentric distance, but it does not
greatly affect to our calculation: 
indeed 60\% of the contribution to the integral stems from $R_1<R<16$ kpc.
On the other hand, the mass which would be lost outwards between $R_0$ and $R_1$ is $\dot{M}(R_0<R<R_1)=-2.0^{+1.6}_{-4.1}$ M$_\odot $/yr. In regions interior to $R<R_0$
there is also mass loss, but this is directed inwards (negative $V_R$).
Therefore, the scenario derived from Eq. (\ref{VRfit}) interpreted as consequence of secular expansion of the disk is that some stars within $R_0<R<R_1$ are migrating to the region $R>R_1$ at the rate of $\sim 2$ M$_\odot $/yr.

The amount of mass of the disk within $R_0<R<R_1$ is 2.2$\times 10^9$ M$_\odot $ and the mass at $R>R_1$ is 1.2$\times 10^9$ M$_\odot $. This means that, with the actual ratio of expansion, the region $R_0<R<R_1$ would
be empty in only 1.1 Gyr and all of the stars within $R>R_1$ would stem from
a migration in the last 0.6 Gyr. This is not possible, since the life of the
Galaxy is much longer and the extension of the Galaxy cannot change so fast.
This means that, either this secular expansion with a motion in an
Archimedes spiral does not apply or these velocities $V_{\rm rm}(R)$ are not
constant with time, so we live now in a period of fast expansion but this period will be short.

The limitations and implications of this scenario are those already mentioned: we are only using the regions toward the center/anticenter, so we can be sure there is a motion of expansion there but not in the rest of the azimuths of the Galaxy. As mentioned above, a local stream cannot be the explanation of radial velocities along a wide range of $\sim 10$ kpc, but it could be a large-scale stream associated with the Galaxy in the Sun-Galactic center line; this idea might have some support in the fact that there
is an asymmetry in the north--south Galactic hemisphere, indicating that this
possible stream would be placed in the southern part. Since we do not have
evidence of such a huge structure embedded in our Galaxy, we think this is not very likely. Assuming the hypothesis of the same radial velocities for any azimuth, as we have done here, we obtain the result of a fast expansion of the disk that would dilute it in few rotations if the expansion were constant with time.
In any case, it may contribute to the increase of the disk size in cosmological times. Most of the mass will be concentrated in the central regions without much change, but the outer disk has a trend at present to lose stars outwards, stars which will escape the Galaxy and will be part of the external halo once they abandon the disc.

\section{Summary and conclusions}

We have obtained a gradient of Galactocentric radial velocities
of $\left \langle \frac{dV_R}{dR}\right \rangle=(+1.48\pm 0.35)$ km/s/kpc (positive indicating outwards with respect to the center of the Galaxy) in the range -3 kpc$<(R-R_\odot )<+8$ kpc,
with a higher contribution to this gradient of the $b<0$ stars.

The point at which the Galactocentric radial velocity is null is at $R_0=8.8\pm 2.7$ kpc.
This is compatible with the mean stellar orbit in the solar Galactocentric radius 
being exactly circular, which is what was also observed by Siebert et al. (2011) and Famaey et al. (2012).

Two possible explanations can be given for this observation:

\begin{enumerate}

\item The mean orbit of the disk stars is intrinsically elliptical in the outer disk.
In such a case, a statistical analysis of our kinematic results gives values of the  eccentricities $e$ for $R>R_0$ with a gradient $\left\langle \frac{de}{dR}\right\rangle \sim 0.01$ kpc$^{-1}$.

\item There is a net secular expansion of the disk associated with migration, in which stars within $R\approx 9-11$ kpc are moving to the region $R\gtrsim 11$ kpc at the rate of $\sim 2$ M$_\odot $/yr. Stars with $R\lesssim 9$ kpc would be falling toward the center of the Galaxy. This expansion would be unattainable for a long time and should be decelerated, otherwise the Galaxy would be fade away
in around 1 Gyr.

\end{enumerate}

We cannot distinguish at present between both scenarios.
At present, both hypotheses are speculative and one would need data on the Galactocentric radial velocity for other azimuths different to the center or anticenter in order to confirm one of the scenarios.

\acknowledgments
The authors are grateful to the anonymous referee for helpful comments.
M.L.C. was supported by the grant AYA2012-33211 of the Spanish Ministry of 
Economy and Competitiveness (MINECO).
This work was supported by the European Science Foundation under the GREAT ESF program, which funded a visit of M.L.C. to the Institute of Astronomy in Cambridge.
SDSS-III is managed by the Astrophysical Research Consortium for the Participating Institutions of the SDSS-III Collaboration including the University of Arizona, the Brazilian Participation Group, Brookhaven National Laboratory, Carnegie Mellon University, the University of Florida, the French Participation Group, the German Participation Group, Harvard University, the Instituto de Astrofisica de Canarias, the Michigan State/Notre Dame/JINA Participation Group, Johns Hopkins University, Lawrence Berkeley National Laboratory, Max Planck Institute for Astrophysics, Max Planck Institute for Extraterrestrial Physics, New Mexico State University, New York University, Ohio State University, Pennsylvania State University, the University of Portsmouth, Princeton University, the Spanish Participation Group, University of Tokyo, the University of Utah, Vanderbilt University, the University of Virginia, the University of Washington, and Yale University.

\end{document}